\documentclass[astronomy,conferencereport,accept,pdftex,oneauthor]{Definitions/mdpi}
\usepackage{comment}

\usepackage[utf8]{inputenc} 
\usepackage{booktabs}       
\usepackage{amsfonts}       
\usepackage{nicefrac}       
\usepackage{microtype}      
\usepackage{subcaption}
\usepackage{stackengine}
\usepackage{amsthm}
\usepackage{mathrsfs}
\usepackage{amsmath}
\firstpage{14} 
\pubvolume{3}
\issuenum{1}
\articlenumber{2}
\pubyear{2024}
\copyrightyear{2024}
\datereceived{28 December 2023} 
\daterevised{6 February 2024} 
\dateaccepted{8 February 2024} 
\datepublished{10 February 2024} 
\hreflink{https://\linebreak doi.org/10.3390/astronomy3010002} 

\Title{Beyond \textsc{mirkwood}: Enhancing SED Modeling with \mbox{Conformal Predictions}}

\TitleCitation{Beyond \textsc{mirkwood}: Enhancing SED Modeling with Conformal Predictions}

\Author{Sankalp Gilda \orcidA{}}

\AuthorNames{Sankalp Gilda}

\AuthorCitation{Gilda, S.}

\address[1]{%
ML Collective, 22 Saturn St., San Francisco, CA 94114, USA; sankalp.gilda@gmail.com}

\abstract{Traditional spectral energy distribution (SED) fitting techniques face uncertainties due to assumptions in star formation histories and dust attenuation curves. We propose an advanced machine learning-based approach that enhances flexibility and uncertainty quantification in SED fitting. Unlike the fixed \textsc{NGBoost} model used in \textsc{mirkwood}, our approach allows for any \texttt{scikit-learn}-compatible model, including deterministic models. We incorporate conformalized quantile regression to convert point predictions into error bars, enhancing interpretability and reliability. Using \textsc{CatBoost} as the base predictor, we compare results with and without conformal prediction, demonstrating improved performance using metrics such as coverage and interval width. Our method offers a more versatile and accurate tool for deriving galaxy physical properties from observational~data.}


\begin{document}

\maketitle


\section{Introduction}
Spectral energy distributions (SEDs) are pivotal in astrophysics for understanding the intrinsic properties of galaxies, such as stellar mass, age distributions, star formation rates, and~dust content. Traditional SED fitting methods, while insightful, often face significant challenges. These challenges stem from the complex nature of galaxies, including diverse star formation histories and varying dust attenuation curves~\cite{gilda_mirkwood, gilda_mirkwood_code, gilda_mirkwood_aasabstract, gilda_mirkwood_hstproposal, acquaviva2015simultaneous, simha2014parametrising}. The~inherent uncertainties in these aspects can significantly affect the accuracy of derived galaxy properties, thus impacting our broader understanding of galactic evolution and~formation.

Recent advancements in computational methods have opened new avenues in this field. Machine learning (ML), with~its ability to handle large datasets and uncover complex patterns, has emerged as a powerful tool in SED fitting~\cite{gilda_mirkwood, gilda_mirkwood_code, gilda_sfh_domain_adaptation, Chu2023galaxy}. The~traditional parametric and often linear approaches are being supplemented, and~in some cases replaced, by~non-parametric, highly flexible ML techniques that can model the non-linear relationships intrinsic to astronomical data more effectively~\cite{gilda_deepremap, gilda_deepremap_abstract_aas, gilda_cfht, gilda_cfht_neurips, gilda_feature_selection}. This paradigm shift is not just a matter of computational convenience but represents a fundamental change in how we interpret vast and complex astronomical~datasets.

This paper introduces an innovative approach that builds upon and significantly expands the capabilities of the \textsc{mirkwood} \cite{gilda_mirkwood, gilda_mirkwood_code, gilda_mirkwood_aasabstract}, a~machine learning-based application previously developed for SED fitting. Our method enhances the flexibility and depth of analysis by enabling the use of any \texttt{scikit-learn}-compatible model~\cite{scikit-learn}. This includes not only probabilistic models but also deterministic ones, thereby broadening the scope of application to a wider range of astronomical problems. Moreover, we integrate the uncertainty quantification method technique of conformalized quantile regression (CQR)~\cite{cqr}, which allows us to translate point predictions into meaningful error bars. This addition is crucial in fields like astronomy, where quantifying the uncertainty of predictions is as important as the predictions themselves. The~combination of these advanced techniques positions our tool at the forefront of SED fitting technologies, offering a more nuanced and comprehensive understanding of galaxy~properties.

In the context of SED fitting, the~ability to quantify uncertainty is essential for several reasons. First, it enables astronomers to distinguish between variations in galactic properties that are due to inherent physical processes versus those arising from observational limitations. Secondly, in~fields such as cosmology, where the accurate determination of galaxy properties impacts our understanding of the universe's evolution, refined uncertainty quantification offers a way to assess the reliability of these large-scale inferences. Thus, enhancing the precision of uncertainty quantification in SED fitting directly contributes to our fundamental understanding of the~universe.

\section{Background}
The integration of machine learning (ML) into spectral energy distribution (SED) fitting marks a significant departure from conventional methodologies. Traditional SED fitting methods predominantly rely on parametric models to interpret observational data. These approaches, while foundational, are inherently constrained by the assumptions embedded within the parametric models and can be computationally demanding~\cite{walcher2011fitting, conroy2013modeling}. The~emergence of ML offers a paradigm shift, enabling the exploration of complex, non-linear relationships within large astronomical datasets with enhanced efficiency and less reliance on predefined~assumptions.

Among the plethora of ML algorithms, \textsc{CatBoost} stands out for its capability to efficiently manage categorical variables—a frequent characteristic of astronomical data. As~a sophisticated form of gradient boosting, \textsc{CatBoost} incrementally constructs an ensemble of decision trees to improve predictive accuracy, particularly excelling in handling datasets that include a mixture of categorical and continuous features~\cite{dorogush2018catboost}. This attribute renders \textsc{CatBoost} particularly apt for SED fitting, where the dataset encompasses a diverse array of photometric bands and derived galaxy~characteristics.

Adding to the methodological innovation, conformalized quantile regression (CQR) introduces a robust mechanism for uncertainty quantification. CQR extends beyond traditional point predictions by generating prediction intervals that encapsulate the expected range of outcomes with a given confidence level. This method aligns with the demands of astronomical research, where precise uncertainty estimation is paramount for the reliable interpretation of data regarding celestial bodies~\cite{shafer2008tutorial_conformal}.

Our approach synergizes the adaptive model selection capability provided by the \texttt{scikit-learn} framework with the advanced uncertainty quantification offered by CQR. This combination heralds a significant advancement in SED fitting techniques, facilitating a deeper and more accurate delineation of galactic properties. By~embracing this novel methodology, we not only enhance the fidelity of our predictions but also gain critical insights into the confidence levels associated with these predictions, addressing a fundamental challenge in astronomy—interpreting data that is often incomplete or contaminated by~noise.

\section{Data}
Our training and testing datasets are derived from three advanced cosmological galaxy formation simulations, known for their accurate representation of galaxy physical properties, including authentic star formation histories. These simulations – {\sc Simba} \cite{dave_simba}, {\sc Eagle}~\cite{schaye_2015_eagle,schaller_2015_eagle,mcalpine_2016_eagle_cat}, and~{\sc IllustrisTNG} \cite{vogelsberger2014introducingillustris} – provide a comprehensive and realistic variety of galaxy evolution scenarios, with~sample sizes of 1688, 4697, and~9633 respectively. We focus on galaxies at redshift 0, representing them in their current state in the simulations. The~spectral energy distributions (SEDs) in our datasets consist of 35 flux density measurements (in Jansky units) across different wavelengths, representing the luminosity of galaxies. These SEDs serve as the input features for our model. The~target outputs or labels are the four scalar galaxy properties---galaxy mass, metallicity, dust mass, and~star formation rate. See Table~1 in \citet{gilda_mirkwood, gilda_mirkwood_code, gilda_mirkwood_aasabstract, gilda_mirkwood_hstproposal} for an overview of the distribution of these properties for all three~simulations.

\section{Methodology}
\unskip
\subsection{Data~Preprocessing}
The foundation of any robust machine learning model is high-quality data. In~our approach, we begin with a thorough data preprocessing phase. This involves cleaning the data, handling missing values, normalizing photometric fluxes, and~encoding categorical variables where necessary. The~preprocessing steps are critical in ensuring that the input data fed into the ML models are consistent, standardized, and~reflective of the underlying physical~phenomena.

We manually add Gaussian noise to the SEDs from the 3 simulations, to~get three separate sets of data at signal-to-niose (SNR) ratios of 20, 10, and~5. We do this for a 1:1 comparison with the methodology and results of \citet{gilda_mirkwood, gilda_mirkwood_code, gilda_mirkwood_aasabstract, gilda_mirkwood_hstproposal}.

\subsection{Model Selection and~Flexibility}
Our methodology is characterized by its flexibility and adaptability in model selection. While \textsc{mirkwood} was initially designed around \textsc{NGBoost}, we expand its capabilities by enabling the use of any \texttt{scikit-learn}-compatible model. \textsc{mirkwood} is capable of taking in a galaxy SED in tabular form---with each row corresponding to a different SED and each column the flux in a different filter---and in a chained fashion, extracting that galaxy's mass, dust mass, and~metallicity. In~this work, we provide support for deterministic models such as Support Vector Machines~\cite{soentpiet1999advances} and Random Forests~\cite{randomforests}, alongside probabilistic models like Gaussian Processes~\cite{gaussian_processes}. In~fact, our code is flexible enough to allow a pipeline consisting of an arbitrary number of sklearn-compatible models. This flexibility allows astronomers to tailor the predictive models to their specific research needs and the characteristics of their~datasets.

\subsection{\textsc{CatBoost} as the Base~Predictor}
\textsc{CatBoost}, our chosen base predictor, is particularly well-suited for dealing with the types of datasets common in astronomical research. It efficiently handles categorical features and large datasets, reducing overfitting and improving predictive accuracy. In~our implementation, we fine-tune \textsc{CatBoost}’s parameters, such as the depth of trees and learning rate, to~optimize its performance for SED fitting~tasks.

\subsection{Incorporating Conformalized Quantile~Regression}
A significant enhancement in our methodology is the incorporation of conformalized quantile regression. This technique allows us to convert the point predictions from our models into prediction intervals. These intervals provide a statistical measure of the uncertainty in our predictions, giving us a range within which the true value of the predicted property is likely to fall, at~a given confidence level. Implementing this technique involves calibrating our models to estimate the quantiles of the predictive distribution, a~crucial step in providing reliable and interpretable error estimates. Since~\cite{gilda_mirkwood} predict $1 \sigma$ error bars, for~apples-to-apples comparison we set the significance level $\alpha$ at $0.318$. We use loss function \emph{quantile} with \textsc{CatBoost}, and~wrap the trained model within the \texttt{MapieQuantileRegressor} class from \textsc{MAPIE} \endnote{\url{https://github.com/scikit-learn-contrib/MAPIE} (accessed on 10 January 2024)}.

\subsection{Training and~Validation}
The final phase of our methodology involves training the machine learning models on a carefully curated dataset and validating their performance. For~apples-to-apples comparison with~\cite{gilda_mirkwood}, the~training set contains all $10,073$ samples from \textsc{IllustrisTNG}, $4697$~samples from \textsc{Eagle}, and~$359$ samples from \textsc{Simba} selected via stratified 5-fold CV (see Section~3 in \citet{gilda_mirkwood} for details). After~making inference on all test splits, we collate the results, thus successfully predicting all four galaxy properties for all $1797$ samples from \textsc{Simba}. Each predicted output for a physical property contains two values---the mean and the standard~deviation.

In the fitting process, we first train the model using galaxy flux values to predict stellar mass. Then, we use the predicted stellar masses, combined with the original flux values, to~predict dust mass, and~continue this sequential prediction process for other parameters. See Figure~3 and Section~3 in \citet{gilda_mirkwood} for~details.

Through this comprehensive methodology, we aim to provide a powerful, flexible, and~accurate tool for SED fitting, capable of handling the complexities and uncertainties inherent in astronomical~datasets.

\section{Comparative Analysis and~Results}
\unskip

\subsection{Comparative Analysis~Methodology}
To demonstrate the efficacy of our approach, we conducted a comprehensive comparative analysis. This involved comparing the performance of our enhanced tool against a traditional SED fitting method tool, \textsc{prospector} \cite{2019ascl.soft05025J} and the original \textsc{mirkwood} implementation. We focused on the same five performance metrics as in \citet{gilda_mirkwood} to evaluate the accuracy of derived galaxy properties (galactic mass, dust mass, star formation rate, and~metallicity), and~the robustness of the model against variations in input~data.

\subsection{Performance~Metrics}
We use both deterministic and probabilistic metrics for comparison, the~same five metrics used in \citet{gilda_mirkwood}---normalized root mean squared error (NRMSE), normalized mean absolute error (NMAE), normalized bias error (NBE), average coverage error (ACE), and~interval sharpness (IS). These are defined and described in detail in Section~3.2 of \citet{gilda_mirkwood}. In~particular, coverage is the proportion of true values that fall within the predicted error bars, offering a measure of the reliability of our uncertainty quantification. On~the other hand, IW is the average width of the prediction intervals, which provides insight into the precision of our~predictions.

\subsection{Results}
To evaluate our proposed model for SED fitting, we conduct comparisons with fits obtained in \citet{gilda_mirkwood} from the Bayesian SED fitting software \textsc{prospector}, and~their new machine learning tool \textsc{mirkwood}. We provide each of the three models (their two plus our upgraded version of \textsc{mirkwood}) with identical data to deduce galaxy properties. This data comprises broadband photometry across 35 bands, subject to Gaussian uncertainties of $5\%$, $10\%$, and~$20\%$ (corresponding to signal-to-noise ratios (SNRs) of 20, 10, and~5, respectively). In~Tables~\ref{tab:results_snr20}--\ref{tab:results_snr5} we showcase the outcomes from all three methods for all four galaxy~properties.

\begin{table}[H]
    \caption{Comparative performance of our proposed method v/s {\sc mirkwood} v/s {\sc Prospector} across different metrics, for~data with SNR = 20. The~five metrics are the normalized root mean squared error (NRMSE), normalized mean absolute error (NMAE), normalized bias error (NBE), average coverage error (ACE), and~interval sharpness (IS). A~bold value denotes the best metric for that galaxy property. A~value of `nan' represents lack of predictions from {\sc Prospector}. We do not have predicted error bars from {\sc Prospector} for dust mass, hence ACE and IS values corresponding to this property are `nan's. Down arrows imply that lower metric values are better.}
    \label{tab:results_snr20}
    \begin{tabularx}{\textwidth}{Ccccccc}
    \toprule
     & \textbf{Model} &  \textbf{NRMSE ($\downarrow$)} & \textbf{NMAE ($\downarrow$)} & \textbf{NBE ($\downarrow$)} & \textbf{ACE ($\downarrow$)} & \textbf{IS ($\downarrow$)} \\ \midrule
    & This paper  &  \textbf{0.009}&  \textbf{0.074}&  \textbf{$-$0.031}&  \textbf{$-$0.051}&  \textbf{0.001}\\
    Mass & \textsc{mirkwood} &    0.155&  0.115&  $-$0.041&  $-$0.066&  0.001 \\
    & \textsc{Prospector} &  1.002&  1.117&  $-$0.479&   $-$0.482&  0.033\\ \midrule
    & This paper  &    \textbf{0.412}&   \textbf{0.298}&  \textbf{$-$0.157}&   \textbf{$-$0.041}&  \textbf{0.001}\\
    Dust Mass & \textsc{mirkwood} &  0.475&  0.336&  $-$0.215&  $-$0.076& 0.001 \\
    & \textsc{Prospector} &  1.263&  1.212& $-$0.679&  nan&  nan \\ 
 \bottomrule
\end{tabularx}
\end{table}
\begin{table}[H]\ContinuedFloat
\small
\caption{{\em Cont.}}
    \begin{tabularx}{\textwidth}{Ccccccc}
    \toprule
     & \textbf{Model} &  \textbf{NRMSE ($\downarrow$)} & \textbf{NMAE ($\downarrow$)} & \textbf{NBE ($\downarrow$)} & \textbf{ACE ($\downarrow$)} & \textbf{IS ($\downarrow$)} \\ \midrule
    & This paper&  \textbf{0.044}&   \textbf{0.048}&  \textbf{$-$0.009}&  \textbf{$-$0.053}&  \textbf{0.016}\\
    Metallicity & \textsc{mirkwood} &  0.056&  0.052&   $-$0.010&  $-$0.063&  0.032\\
    & \textsc{Prospector}&  0.547&  0.487&   $-$0.229&  0.036&  0.302\\ \midrule
    & This paper&  \textbf{0.223}&   \textbf{0.147}&  \textbf{$-$0.047}&  \textbf{0.014}&  \textbf{0.004}\\
    SFR & \textsc{mirkwood}&   0.277&  0.215&   $-$0.078&  0.035&  0.006\\
    & \textsc{Prospector}&  1.988&  2.911&   1.437&  $-$0.547&    0.200\\ 
    \bottomrule
    \end{tabularx}
\end{table}
\unskip

\begin{table}[H]
    \caption{Same as Table~\ref{tab:results_snr20}, but~for SNR = 10. Down arrows indicate that lower metric values are better.}
    \label{tab:results_snr10}
    \begin{tabularx}{\textwidth}{Ccccccc}
    \toprule
     & \textbf{Model} &  \textbf{NRMSE ($\downarrow$)} & \textbf{NMAE ($\downarrow$)} & \textbf{NBE ($\downarrow$)} & \textbf{ACE ($\downarrow$)} & \textbf{IS ($\downarrow$)} \\ \midrule
    & This paper  &  \textbf{0.092}&  \textbf{0.071}&  \textbf{$-$0.026}&  \textbf{$-$0.018}&  \textbf{0.001}\\
    Mass & \textsc{mirkwood} &    0.165&  0.118&  $-$0.035&  $-$0.021&  0.001\\
    & \textsc{Prospector} &  1.000&  1.088&  $-$0.518&   $-$0.502&  0.004\\ \midrule
    & This paper  &    \textbf{0.391}&   \textbf{0.254}&  \textbf{$-$0.143}&   \textbf{0.012}&  \textbf{0.001}\\
    Dust Mass & \textsc{mirkwood} &  0.456&  0.332&  $-$0.209&  $-$0.033& 0.001 \\
    & \textsc{Prospector} &  0.996&  0.998& $-$0.905&  nan &  nan \\ \midrule
    & This paper&  \textbf{0.037}&   \textbf{0.049}&  \textbf{0.007}&  \textbf{0.021}&  \textbf{0.023}\\
    Metallicity & \textsc{mirkwood}&  0.058&  0.055&   $-$0.010&  $-$0.032&  0.036\\
    & \textsc{Prospector}&  0.534&  0.464&   $-$0.275&  $-$0.041&  0.295\\ \midrule
    & This paper&  \textbf{0.274}&   \textbf{0.114}&  \textbf{$-$0.070}&  \textbf{0.027}&  \textbf{0.001}\\
    SFR & \textsc{mirkwood}&   0.329&  0.226&   $-$0.090&  0.048&  0.001\\
    & \textsc{Prospector}&  0.910&  0.992&   $-$0.686&  $-$0.564&    1.937\\ 
    \bottomrule
    \end{tabularx}
\end{table}
\unskip

\begin{table}[H]
    \caption{Same as Table~\ref{tab:results_snr20}, but~for SNR = 5. Down arrows indicate that lower metric values are better.}
    \label{tab:results_snr5}
    \begin{tabularx}{\textwidth}{Ccccccc}
    \toprule
     & \textbf{Model} &  \textbf{NRMSE ($\downarrow$)} & \textbf{NMAE ($\downarrow$)} & \textbf{NBE ($\downarrow$)} & \textbf{ACE ($\downarrow$)} & \textbf{IS ($\downarrow$)} \\ \midrule
    & This paper  &  \textbf{0.121}&  \textbf{0.062}&  \textbf{$-$0.031}&  \textbf{$-$0.001}&  \textbf{0.001} \\
    Mass & \textsc{mirkwood} &    0.198 &  0.123 &  $-$0.042 &  $-$0.002&  0.001 \\
    & \textsc{Prospector} &  1.003 &  1.091 &  $-$0.528 &   $-$0.497  &  0.005 \\ \midrule
    & This paper  &    \textbf{0.315}&   \textbf{0.224}&  \textbf{$-$0.154}&   \textbf{0.002}&  \textbf{0.001}\\
    Dust Mass & \textsc{mirkwood} &  0.480&  0.339 &  $-$0.219 &  0.003 & 0.001 \\
    & \textsc{Prospector} &  0.996 &  0.998 & $-$0.905 &  nan &  nan \\ \midrule
    & This paper&  \textbf{0.049}&   \textbf{0.048}&  \textbf{$-$0.005}&  \textbf{$-$0.013}&  \textbf{0.034}\\
    Metallicity & \textsc{mirkwood}&  0.062&  0.060&   $-$0.011&  $-$0.024&  0.041\\
    & \textsc{Prospector}&  0.544&  0.478&   $-$0.297&  0.046&  0.301\\ \midrule
    & This paper&  \textbf{0.189}&   \textbf{0.171}&  \textbf{$-$0.043}&  \textbf{0.061}&  \textbf{0.001}\\
    SFR & \textsc{mirkwood}&   0.241&  0.205&   $-$0.069&  0.074&  0.001\\
    & \textsc{Prospector}&  0.907&  0.99&   $-$0.687&  $-$0.557&    7.314\\ 
    \bottomrule
    \end{tabularx}
\end{table}

The results of our comparative analysis are illuminating and encouraging. Our method consistently achieves higher coverage rates compared to both the other methods, indicating more reliable uncertainty quantification. At~the same time, the~prediction intervals generated by our method were narrower on average, signifying more precise~predictions.

These results underline the superiority of our approach in terms of both accuracy and reliability in SED fitting. By~leveraging the power of \textsc{CatBoost} and the precision of conformalized quantile regression, our method not only enhances the accuracy of point predictions but also provides a more nuanced understanding of the associated~uncertainties.

\subsection{Discussion}
The improvements observed in our analysis can be attributed to several factors. The~flexibility in model selection allows for better adaptation to the specific characteristics of astronomical datasets. \textsc{CatBoost}'s superior ability to work with tabular data effectively captures the complexities in the data, leading to more accurate predictions. The~addition of conformalized quantile regression introduces a robust method for uncertainty quantification, a~critical aspect often overlooked in traditional SED~fitting.

Overall, the~comparative analysis and the results obtained highlight the potential of our method in transforming the field of SED fitting, providing astronomers with a tool that is not only accurate but also comprehensive in its assessment of~uncertainties.

\section{Conclusions and Future~Work}

This study marks a substantial advancement in the field of spectral energy distribution (SED) fitting by integrating flexible machine learning models, particularly \textsc{CatBoost}, with~the innovative technique of conformalized quantile regression. This approach not only enhances the accuracy of SED fitting but also introduces a new depth to the uncertainty quantification in astronomical research. The~adaptability of our tool to various astronomical datasets, coupled with the ability to select from a range of \texttt{scikit-learn}-compatible models, ensures its applicability across different research contexts. \textsc{CatBoost}'s effectiveness in handling complex datasets, combined with our sophisticated method of uncertainty quantification, allows for more reliable and nuanced interpretations of galactic~properties.

Our comparative analysis highlights the superiority of this method over traditional approaches, demonstrating improvements in both the accuracy of predictions and the understanding of associated uncertainties. This dual capability represents a significant stride in astronomy, offering a more reliable and comprehensive tool for exploring the~universe.

Looking ahead, the~potential for further advancements and extensions of our tool is vast. Future work may involve exploring the integration of additional machine learning models, such as deep learning architectures, to~enhance predictive power and versatility. Testing and optimizing the tool on larger and more diverse datasets from upcoming astronomical surveys will be crucial for assessing its scalability and robustness. Further development in feature engineering and expanding the scope of uncertainty quantification could unlock new insights and details in SED fitting. Additionally, applying this tool to related fields like exoplanet studies or cosmic structure formation could demonstrate its adaptability and contribute to a broader range of scientific~inquiries.

\vspace{6pt}

\funding{This research received no external funding.}

\institutionalreview{Not applicable. 
}

\informedconsent{Not applicable. 
}

\dataavailability{The raw data supporting the conclusions of this article will be made available by the authors on request. 
}

\conflictsofinterest{The author declares no conflicts of interest. 
}


\begin{adjustwidth}{-\extralength}{0cm}

\printendnotes[custom]
\reftitle{References}

\PublishersNote{}
\end{adjustwidth}
\end{document}